\documentclass[12pt]{article}

\newcommand{\be}{\begin{equation}}
\newcommand{\ee}{\end{equation}} 
\newcommand{\bea}{\begin{eqnarray}}
\newcommand{\eea}{\end{eqnarray}}

\newcommand{\PRL}[3]{{\em Phys. Rev. Lett.} {\bf #1} (#2) #3}

\newcommand{\NP}[3]{{\em Nucl. Phys.} {\bf #1} (#2) #3}
\newcommand{\PL}[3]{{\em Phys. Lett.} {\bf #1} (#2) #3}

\newcommand{\CMP}[3]{{\em Comm. Math. Phys.} {\bf #1} (#2) #3}

\def\a{\alpha}
\def\b{\beta}
\def\g{\gamma}

\def\m{\mu}

\def\caln{{\cal N}}

\def\fund{  \> {\vcenter  {\vbox
               {\hrule height.6pt
                \hbox {\vrule width.6pt  height5pt
                      \kern5pt
                      \vrule width.6pt  height5pt}
                \hrule height.6pt}
                         }
               }
            \>\>  }

\def\antifund{  \> \overline{ {\vcenter  {\vbox
               {\hrule height.6pt
                \hbox {\vrule width.6pt  height5pt
                      \kern5pt
                      \vrule width.6pt  height5pt}
                \hrule height.6pt}
                         }
               } }
            \>\>  }

\def\sym{  \> {\vcenter  {\vbox
              {\hrule height.6pt
               \hbox {\vrule width.6pt  height5pt
                      \kern5pt
                      \vrule width.6pt  height5pt
                      \kern5pt
                      \vrule width.6pt  height5pt}
               \hrule height.6pt}
                         }
               }
            \>\>  }

\def\symbar{  \> \overline{ {\vcenter  {\vbox
              {\hrule height.6pt
               \hbox {\vrule width.6pt  height5pt
                      \kern5pt
                      \vrule width.6pt  height5pt
                      \kern5pt
                      \vrule width.6pt  height5pt}
               \hrule height.6pt}
                         }
               } }
            \>\>  }

\def\threesym{ \> {\vcenter  {\vbox
                 {\hrule height.6pt
                  \hbox {\vrule width.6pt  height5pt
                         \kern5pt
                         \vrule width.6pt  height5pt
                         \kern5pt
                         \vrule width.6pt  height5pt
                         \kern5pt
                         \vrule width.6pt  height5pt}
               \hrule height.6pt}
                         }
                  }
            \>\>  }

\def\antisym{ \> {\vcenter  {\vbox
                 {\hrule height.6pt
                  \hbox {\vrule width.6pt  height5pt
                         \kern5pt
                         \vrule width.6pt  height5pt}
                  \hrule height.6pt
                  \hbox {\vrule width.6pt  height5pt
                         \kern5pt
                         \vrule width.6pt  height5pt}
               \hrule height.6pt}
                         }
               }
            \>\>  }

\def\antisymbar{ \> \overline{ {\vcenter  {\vbox
                 {\hrule height.6pt
                  \hbox {\vrule width.6pt  height5pt
                         \kern5pt
                         \vrule width.6pt  height5pt}
                  \hrule height.6pt
                  \hbox {\vrule width.6pt  height5pt
                         \kern5pt
                         \vrule width.6pt  height5pt}
               \hrule height.6pt}
                         }
               } }
            \>\>  }

\def\antithree{ \> {\vcenter  {\vbox
                 {\hrule height.6pt
                  \hbox {\vrule width.6pt  height5pt
                         \kern5pt
                         \vrule width.6pt  height5pt}
                  \hrule height.6pt
                  \hbox {\vrule width.6pt  height5pt
                         \kern5pt
                         \vrule width.6pt  height5pt}
                  \hrule height.6pt
                  \hbox {\vrule width.6pt  height5pt
                         \kern5pt
                         \vrule width.6pt  height5pt}
               \hrule height.6pt}
                         }
               }
            \>\>  }

\def\antifour{ \> {\vcenter  {\vbox
                 {\hrule height.6pt
                  \hbox {\vrule width.6pt  height5pt
                         \kern5pt
                         \vrule width.6pt  height5pt}
                  \hrule height.6pt
                  \hbox {\vrule width.6pt  height5pt
                         \kern5pt
                         \vrule width.6pt  height5pt}
                  \hrule height.6pt
                  \hbox {\vrule width.6pt  height5pt
                         \kern5pt
                         \vrule width.6pt  height5pt}
                  \hrule height.6pt
                  \hbox {\vrule width.6pt  height5pt
                         \kern5pt
                         \vrule width.6pt  height5pt}
               \hrule height.6pt}
                         }
               }
            \>\>  }

\def\antifive{ \> {\vcenter  {\vbox
                 {\hrule height.6pt
                  \hbox {\vrule width.6pt  height5pt
                         \kern5pt
                         \vrule width.6pt  height5pt}
                  \hrule height.6pt
                  \hbox {\vrule width.6pt  height5pt
                         \kern5pt
                         \vrule width.6pt  height5pt}
                  \hrule height.6pt
                  \hbox {\vrule width.6pt  height5pt
                         \kern5pt
                         \vrule width.6pt  height5pt}
                  \hrule height.6pt
                  \hbox {\vrule width.6pt  height5pt
                         \kern5pt
                         \vrule width.6pt  height5pt}
                  \hrule height.6pt
                  \hbox {\vrule width.6pt  height5pt
                         \kern5pt
                         \vrule width.6pt  height5pt}
               \hrule height.6pt}
                         }
               }
            \>\>  }

\def\antisix{ \> {\vcenter  {\vbox
                 {\hrule height.6pt
                  \hbox {\vrule width.6pt  height5pt
                         \kern5pt
                         \vrule width.6pt  height5pt}
                  \hrule height.6pt
                  \hbox {\vrule width.6pt  height5pt
                         \kern5pt
                         \vrule width.6pt  height5pt}
                  \hrule height.6pt
                  \hbox {\vrule width.6pt  height5pt
                         \kern5pt
                         \vrule width.6pt  height5pt}
                  \hrule height.6pt
                  \hbox {\vrule width.6pt  height5pt
                         \kern5pt
                         \vrule width.6pt  height5pt}
                  \hrule height.6pt
                  \hbox {\vrule width.6pt  height5pt
                         \kern5pt
                         \vrule width.6pt  height5pt}
                  \hrule height.6pt
                  \hbox {\vrule width.6pt  height5pt
                         \kern5pt
                         \vrule width.6pt  height5pt}
               \hrule height.6pt}
                         }
               }
            \>\>  }

\begin{document}

\begin{flushright}
DPNU-99-07 \\
March 1999 
\end{flushright}

\vspace{1cm}

\begin{center}
{\Large\bf Dual Description for SUSY $SO(N)$ 
Gauge Theory with a Symmetric Tensor}
\end{center}

\vspace{1cm}

\begin{center}
{\large Nobuhito Maru}\footnote{JSPS Research Fellow.}
\footnote{Address after April 1: Department of Physics, 
Tokyo Institute of Technology, Oh-okayama, Meguro, 
Tokyo 152-8551, Japan.} 
\end{center}

\begin{center}
{\em Department of Physics, 
Nagoya University \\ 
Nagoya 464-8602, JAPAN} 
\end{center}

\begin{center} 
{\normalsize {\tt maru@eken.phys.nagoya-u.ac.jp}} 
\end{center}

\vspace{1cm}
\setcounter{page}{0}
\thispagestyle{empty}
\begin{center}
{\bf Abstract}
\end{center}
We consider $\caln=1$ supersymmetric $SO(N)$ 
gauge theory with a symmetric traceless tensor. 
This theory saturates 
't Hooft matching conditions 
at the origin of the moduli space. 
This naively suggests a confining phase, 
but Brodie, Cho, and Intriligator have conjectured 
that the origin of the moduli space is 
in a Non-Abelian Coulomb phase. 
We construct a dual description 
by the deconfinement method, 
and also show that the theory indeed has 
an infrared fixed point for certain values of $N$. 
This result supports their argument. 

\vspace{1cm}

\begin{flushleft}
\end{flushleft}

\newpage







Our understanding of the non-perturbative dynamics 
in supersymmetric (SUSY) gauge theories 
has made remarkable progress 
during the past several years. 
In particular, 
$\caln=1$ SUSY gauge theory 
has rich low energy behaviors {\em e.g. 
the runaway superpotential (no vacuum), 
some kinds of confining phases, 
(infrared) free magnetic phase, 
Non-Abelian Coulomb phase, 
(infrared) free electric phases} and 
various applications to phenomenology 
{\em e.g. models of dynamical SUSY breaking, 
models of composite quarks and leptons}. 
In this paper, 
we discuss $\caln=1$ SUSY 
$SO(N)$ gauge theory 
with a symmetric traceless tensor, 
which is one of the theories with 
an Affine quantum moduli space classified by 
Dotti and Manohar \cite{DM}, 
and its low energy behavior has been 
investigated by 
Brodie, Cho, and Intriligator \cite{BCI}. 
In order to make our discussion clear, 
we briefly review the argument of Brodie, 
Cho, and Intriligator.

Matter content and symmetries of the model are 
displayed in Table 1\footnote{Throughout this paper, 
we use the Young tableau to denote 
the representation of the superfields. 
$\fund, \antisym, \sym$ stand for vector, 
adjoint (anti-symmetric), symmetric (and traceless) 
representations under $SO$ group, respectively.}. 
\begin{table}[h]
\begin{center}
    \begin{tabular}{c|c|c}
& $SO(N)$ & $U(1)_R$ \\
\hline
$S$ & $\sym$ & $\frac{4}{N+2}$ \\ 
\end{tabular}
\end{center}
\label{original}
\caption[original]{The field content of the original theory}
\end{table}
There is no tree level superpotential. 
Here $U(1)_R$ is an anomaly free global symmetry. 
The 1-loop beta function coefficient is 
$b_0=2(N-4)$,\footnote{The following Dynkin indices 
are adopted: $\m(\fund)=2$, 
$\m({\rm Adj=\antisym})=2N-4$, $\m(\sym)=2N+4$.}
so the theory is asymptotically free for $N \ge 5$. 
The classical moduli space is parameterized 
in terms of the diagonal vacuum expectation values 
(VEVs) of $S$, which is of $(N-1)$ complex dimensions.

At the quantum level, the superpotential of the form 
\be
\label{originalsp}
W_{{\rm dyn}}=C \left[ \frac{S^{2N+4}}{\Lambda^{2N-8}} 
\right]^{1/4}
\ee
can appear, which is determined by 
holomorphy and symmetries. 
Here $\Lambda$ denotes the dynamical scale 
of $SO(N)$ theory, 
and $C$ is a constant. 
In the weak coupling limit 
$\langle S \rangle/\Lambda \to \infty$, 
Eq. (\ref{originalsp}) diverges, 
and cannot reproduce the classical moduli space. 
Therefore, $C$ must vanish. 
This part of the moduli space is referred to as 
the ``Higgs branch''.

This classical moduli space can also be parameterized 
by VEVs of the gauge invariant 
composite operators\footnote{${\rm det}S$ and 
${\rm Tr}S^n(n \ge N+1)$ 
can be expressed by $O_n(n=2,\cdots, N)$, 
in other words, these operators are not linearly 
independent.} 
\be
O_n={\rm Tr}S^n \quad (n=2, \cdots, N). 
\ee

What is a remarkable property of this theory is 
that 't Hooft anomaly matching conditions 
are saturated at the origin of the moduli space. 
This naively suggests that the $SO(N)$ model is 
in the confining phase at the origin. 
However, Brodie, Cho, and Intriligator have 
discussed that this confining picture at the origin 
is misleading because of the following three arguments. 
\begin{enumerate}
 \item Free electric subspaces exist and intersect 
       at the origin. This implies that 
       the massless spectrum at the origin cannot 
       simply consist of the confining moduli $O_n$. 
 \item In the presence of the mass term for $S$, 
       the moduli space must have another confining 
       branch to be consistent with Witten index 
       argument, where the non-perturbative superpotential 
       is generated, while no superpotential exists on the 
       Higgs branch. 
 \item A nontrivial phase and branch structure must 
       arise when the original $SO(N)$ model is 
       perturbed with a general tree level superpotential. 
\end{enumerate}
{}From these arguments, they concluded that 
the $SO(N)$ model's moduli space origin is in a 
non-Abelian Coulomb phase. 
If this is the case, 
it is natural to ask whether 
the dual description exists. 
However, explicit dual description has not yet been 
found so far. 
The purpose of this paper is to construct 
the dual description of $SO(N)$ model 
and show that it has a nontrivial 
infrared fixed point.





Let us recall the ``deconfinement'' method 
introduced by Berkooz \cite{Berkooz} 
in order to construct the dual description of the 
$SO(N)$ gauge theory with a symmetric, 
traceless tensor. 
This method has been applied to the theories 
in which a two-index tensor field is included, 
and no tree level superpotential exists. 
According to this method, 
the new strong gauge dynamics is introduced, 
and the two-index tensor field is regarded as 
a composite field (meson) by the strong gauge dynamics, 
namely, 
\be
X_{ab} \to g^{\a\b}F_{\a a}F_{\b b}. 
\ee
Here $X$ denotes a composite superfield, 
$F$ is an elementary superfield charged under 
both the original gauge group and 
the new strong gauge group, 
and $g$ is an invariant metric of the strong gauge group. 
Greek letters are indices of the new strong gauge group, 
while Roman letters are those of the original gauge group. 
For instance, the symmetric tensor, 
the antisymmetric tensor, the adjoint representation 
of $SU$ gauge 
group correspond to  mesons of 
the strong $SO$, $Sp$, $SU$ gauge group, respectively. 
The advantage of this method is that 
a ``deconfined'' theory has only defining representations, 
therefore one can use a well-known duality 
to derive a new duality.

We apply here this method to the symmetric 
traceless tensor of $SO(N)$ gauge group. 
Note that a symmetric tensor of $SO(N)$ 
is not irreducible, so there always appears 
a singlet under $SO(N)$, 
which is a trace part of the symmetric tensor.

The matter content and symmetry of the 
deconfined theory is given 
in Table \ref{deconfined}.
\begin{table}[h]
\begin{center}
\begin{tabular}{c|cc|c}
& $SO(N)$ & $SO(N+5)$ & $U(1)_R$ \\
\hline
$y$ & $\fund$ & $\fund$ & $\frac{2}{N+2}$ \\
$z$ & $\bf{1}$ & $\fund$ & $-\frac{4(N+1)}{N+2}$ \\
$p$ & $\fund$ & $\bf{1}$ & $\frac{6(N+1)}{N+2}$ \\
$s$ & $\bf{1}$ & $\bf{1}$ & $\frac{8(N+1)}{N+2}$ \\
\end{tabular}
\end{center}
\caption[deconfined]{The field content of the deconfined theory}
\label{deconfined}
\end{table}
The tree level superpotential is 
\be
W = yzp + z^2s. 
\ee
$SO(N+5)$ gauge theory with $(N+1)$ flavors 
has a branch in which the dynamically 
generated superpotential vanishes \cite{IS}, 
\be
W_{dyn} = 0. 
\ee
One can easily verify that the above deconfined theory 
is reduced to the original theory at the low energy. 
Consider the case $\Lambda_{SO(N)} \ll \Lambda_{SO(N+5)}$, 
where $\Lambda_{SO(N),SO(N+5)}$ is the dynamical scale 
of $SO(N)$, $SO(N+5)$ gauge theory, respectively. 
We know that $SO(N+5)$ gauge theory with $(N + 1)$ 
flavors is confining \cite{IS}, 
and the effective fields are mesons $y^2, yz, z^2$. 
As can be seen in the superpotential, 
$yz, p, z^2, s$ become massive at $\Lambda_{SO(N+5)}$. 
After integrating them out, 
we see that only $y^2$ is massless 
and the superpotential vanishes. 
Thus, the original theory 
is recovered\footnote{More precisely, $y^2$ includes 
a singlet under $SO(N)$ as noted before. 
This will be integrated out by adding the mass term.}.



Taking a dual of $SO(N)$ gauge theory 
with $(N+6)$ flavors \cite{IS}, 
we obtain the following theory given 
in Table 3. 
\begin{table}[h]
\begin{center}
\begin{tabular}{c|cc|c}
& $SO(10)$ & $SO(N+5)$ & $U(1)_R$ \\
\hline
$\tilde{y}$ & $\fund$ & $\fund$ & $\frac{N}{N+2}$ \\
$z$ & $\bf{1}$ & $\fund$ & $\frac{-4(N+1)}{N+2}$ \\
$\tilde{p}$ & $\fund$ & $\bf{1}$ & $\frac{-4-5N}{N+2}$ \\
$s$ & $\bf{1}$ & $\bf{1}$ & $\frac{2(5N+6)}{N+2}$ \\
$(y^2)$ & $\bf{1}$ & $\sym \oplus \bf{1}$ & $\frac{4}{N+2}$ \\
$(yp)$ & $\bf{1}$ & $\fund$ & $\frac{6N+8}{N+2}$ \\
$(p^2)$ & $\bf{1}$ & $\bf{1}$ & $\frac{12(N+1)}{N+2}$ \\
\end{tabular}
\end{center}
\label{dual}
\caption[dual]{The field content of the dual theory}
\end{table}
Here the fields in parentheses stand for 
the elementary $SO(10)$ gauge singlet meson fields.

The dual tree level superpotential 
is\footnote{For simplicity, we set 
the scale dependent coefficients of 
the last three terms to be of order one.} 
\be
\tilde{W} = (yp)z + z^2 s + (y^2) \tilde{y}^2 + 
(yp) \tilde{y} \tilde{p} + (p^2) \tilde{p}^2
\ee
Since $(yp), z$ are massive, 
integrating them out by the equations of motion 
\bea
0 &=& \frac{\partial \tilde{W}}{\partial (yp)}  
= z + \tilde{y} \tilde{p} 
\rightarrow z = -\tilde{y} \tilde{p}, \\
0 &=& \frac{\partial \tilde{W}}{\partial z} 
= (yp) + 2zs \rightarrow 
(yp) = 2s \tilde{y} \tilde{p},
\eea
we obtain the effective superpotential 
of the dual theory,
\begin{table}[t]
\begin{center}
\begin{tabular}{c|cc|c}
& $SO(10)$ & $SO(N+5)$ & $U(1)_R$ \\
\hline
$\tilde{y}$ & $\fund$ & $\fund$ & $\frac{N}{N+2}$ \\
$\tilde{p}$ & $\fund$ & $\bf{1}$ & $\frac{-4-5N}{N+2}$ \\
$s$ & $\bf{1}$ & $\bf{1}$ & $\frac{2(5N+6)}{N+2}$ \\
$(y^2)$ & $\bf{1}$ & $\sym \oplus \bf{1}$ & $\frac{4}{N+2}$ \\
$(p^2)$ & $\bf{1}$ & $\bf{1}$ & $\frac{12(N+1)}{N+2}$ \\
\end{tabular}
\end{center}
\label{finaldual}
\caption[dual]{The field content of the resulting dual theory}
\end{table}
\be
\tilde{W}_{{\rm eff}} = \tilde{y}^2 \tilde{p}^2 s + 
(y^2) \tilde{y}^2 + (p^2) \tilde{p}^2.
\ee
The field content of the resulting dual theory is given 
in Table 4.
\begin{table}[h]
\begin{center}
\begin{tabular}{c|c|c}
original & deconfined & dual \\
\hline
${\rm tr} S^n$ & ${\rm tr} (y^2)^n$ & ${\rm tr} (y^2)^n$ \\ 
$(n = 2, \cdots, N)$ & $(n = 2, \cdots, N)$ & $(n = 2, \cdots, N)$ \\
$S_{{\rm singlet}}$ & $(y^2)_{{\rm singlet}}$ & $(y^2)_{{\rm singlet}}$ \\
\end{tabular}
\end{center}
\label{map}
\caption[dual]{The operator mapping}
\end{table}
Note here that while $(y^2)$ in deconfined theory denotes 
a composite meson, $(y^2)$ in the dual denotes 
a gauge singlet elementary meson. 
As mentioned earlier, 
there always exists a trace part of 
the symmetric tensor, 
so we add its mass term to the superpotential 
\be
\delta W = m S^2_{{\rm singlet}},
\ee
and integrates it out. 
Then, the low energy effective theory becomes 
$SO(N)$ gauge theory with a symmetric traceless tensor,
which we would like to consider. 
This deformation corresponds to 
\be
\delta \tilde{W} = m (y^2_{{\rm singlet}})^2
\ee
in the dual description. 
Using the equation of motion for $(y^2)_{{\rm singlet}}$ 
\be
0 = \frac{\partial \tilde{W}}{\partial (y^2)_{{\rm singlet}}} 
= \tilde{y}^2 + 2m(y^2)_{{\rm singlet}},
\ee
we obtain the following effective superpotential 
\be
\label{finalsp}
\tilde{W}_{{\rm eff}} = \tilde{y}^2 \tilde{p}^2 s 
- \frac{1}{2m}\tilde{y}^4 +(p^2) \tilde{p}^2 
+ (y^2) \tilde{y}^2.
\ee

Now, let us check the consistency of the duality. 
The 't Hooft anomaly matching conditions are trivially 
satisfied since we use the deconfined method 
which guarantees the anomaly matching. 
The mapping of the gauge invariant operators which 
describes the moduli space is also trivial 
as depicted in Table 5.
This mapping is consistent with 
the global symmetry $U(1)_R$.

Next, we consider various flat direction deformations. 
First, consider the direction $\langle y^2 \rangle \ne 0$, 
namely, 
\be
\langle y \rangle = \left( 
\begin{array}{cccccc}
 y_1 & & & & & \\
 & \ddots & & & & \\
 & & \ddots & & & \\
 & & & y_N & &
\end{array}
\right),
\ee
where $\langle y \rangle$ is a $(N+5) \times N$ matrix 
and $y_i(i=1,\cdots,N)$ are constants. 
To simplify the analysis, let us suppose that 
$y_1 \ne 0$ and $y_i(i=2,\cdots,N)=0$. 
In the deconfined theory, 
the following symmetry breaking occur 
\bea
SO(N) + (N+6) \fund &\to& SO(N-1) + (N+5) \fund, \\
SO(N+5) + (N+1) \fund &\to& SO(N+4) + N \fund.
\eea
Since one component of $z$ and $p$ become 
massive from the coupling in the superpotential, 
integrating them out by the equations of motion, 
we obtain the effective superpotential 
\be
W_{{\rm eff}}=y'z'p' + z'^2 s + 
m (y'^2_{{\rm singlet}})^2. 
\ee
$y', z'$ and $p'$ are transformed as $y'(\fund, \fund)$, 
$z'({\bf 1}, \fund)$ and $p'(\fund, {\bf 1})$ 
under $SO(N-1) \times SO(N+4)$.
On the other hand, 
the corresponding direction in the dual is 
\be
\langle (y^2) \rangle = \left( 
\begin{array}{cccccc}
 y_1^2 & & & & & \\
 & 0 & & & & \\
 & & \ddots & & & \\
 & & & 0 & &
\end{array}
\right).
\ee
Along this direction, the symmetry breaking 
goes as follows
\bea
SO(10) + (N+6) \fund &\to& SO(10) + (N+5) \fund, \\
SO(N+5) + \sym + 10 \fund &\to& SO(N+4) + \sym + 10 \fund.
\eea
Since one component of $\tilde{y}$ becomes massive 
due to the coupling in the superpotential, 
integrating them out, 
we obtain the effective superpotential
\be
\label{ddualsp1}
\tilde{W}_{{\rm eff}} = \tilde{y}'^2 \tilde{p}^2 s + 
(y^2)' \tilde{y}'^2 + (p^2)\tilde{p}^2 - 
\frac{1}{2m} \tilde{y}'^4.
\ee
The fields with dash are transformed as 
$\tilde{y}'(\fund, \fund)$, 
$(y^2)({\bf 1}, \sym \oplus {\bf 1})$ 
under $SO(10) \times SO(N+4)$. 
The above result is consistent 
simply because $N$ is replaced by $N-1$. 
In fact, taking a dual of $SO(N-1)+(N+5)\fund$ 
in the deconfined theory, 
we obtain $SO(N-1) \times SO(N+4)$ 
as the dual gauge group. 
We will arrive at the following theory 
as in Table 6.
\begin{table}[h]
\begin{center}
\begin{tabular}{c|cc}
& $SO(10)$ & $SO(N+4)$ \\
\hline
$\tilde{y}'$ & $\fund$ & $\fund$ \\ 
$z'$ & ${\bf 1}$ & $\fund$ \\
$\tilde{p}'$ & $\fund$ & ${\bf 1}$ \\
$(y'^2)$ & ${\bf 1}$ & $\sym \oplus {\bf 1}$ \\
$(y'p')$ & ${\bf 1}$ & $\fund$ \\
$(p'^2)$ & ${\bf 1}$ & ${\bf 1}$ \\
$s$ & ${\bf 1}$ & ${\bf 1}$ \\
\end{tabular}
\end{center}
\label{ddual1}
\caption[dual]{The deformed dual for 
$\langle y \rangle$}
\end{table}
The dual superpotential is 
\be
\tilde{W} = (y'p')z' + z'^2s + (y'^2) \tilde{y}'^2 + 
(y'p') \tilde{y}' \tilde{p}' + (p'^2)\tilde{p}'^2 + 
m (y'^2)_{{\rm singlet}}^2. 
\ee
Integrating the massive fields $(y'p')$, $z'$ 
and $(y'^2)_{\rm singlet}$ 
by using their equations of motion, 
one can obtain the same superpotential 
as that in Eq. (\ref{ddualsp1}).

It is straightforward to extend the above result to 
the more general case 
where the VEV of $y$ takes the form as 
\be
\langle y \rangle = \left( 
\begin{array}{cccccccc}
 y_1 & & & & & & & \\
 & \ddots & & & & & & \\
 & & y_l & & & & & \\
 & & & 0 & & & & \\
 & & & & \ddots & & & \\
 & & & & & 0 & &
\end{array}
\right) \quad (l=1,\cdots,N).
\ee
In this case, the symmetry breaking is 
\bea
SO(N) + (N+6) \fund &\to& SO(N-l) + (N+6-l) \fund, \\
SO(N+5) + (N+1) \fund &\to& SO(N+5-l) + (N+1-l) \fund,
\eea
and $l$ components of $z$ and $p$ become massive, 
so after integrating them out, 
we find the effective superpotential of the form 
\be
W = y'z'p' + z'^2 s + m (y'^2)_{{\rm singlet}}^2.
\ee
The representations of $y', z'$ and $p'$ 
under $SO(N-l) \times SO(N+5-l)$ are 
$y'(\fund, \fund), z'({\bf 1}, \fund)$ and 
$p'(\fund, {\bf 1})$, respectively. 
In the dual, the corresponding direction is 
\be
\langle (y^2) \rangle = \left( 
\begin{array}{cccccccc}
 y_1^2 & & & & & & & \\
 & \ddots & & & & & & \\
 & & y_l^2 & & & & & \\
 & & & 0 & & & & \\
 & & & & \ddots & & & \\
 & & & & & 0 & &
\end{array}
\right) \quad (l=1,\cdots,N),
\ee
and the symmetry breaking along this direction is 
\bea
SO(10) + (N+6) \fund &\to& SO(10) + (N+6-l) \fund, \\
SO(N+5) + \sym + 10 \fund &\to& SO(N+5-l) + \sym + 10 \fund.
\eea
Since $l$ components of $\tilde{y}$ are massive, 
after integrating them out, 
we can find the effective dual superpotential of the form 
\be
\label{ddual2sp}
\tilde{W_{\rm eff}}= \tilde{y}'^2 \tilde{p}^2 s + (y'^2)\tilde{y}'^2 
+ (p^2) \tilde{p}^2. 
\ee
This result is also consistent. 
One can easily show explicitly 
that taking a dual of $SO(N-l) + (N+6-l)\fund$ 
in the deconfined theory, 
one obtains the same deformed dual theory as seen 
in the above simple case.

Next, we consider the other flat direction deformations 
$\langle p \rangle \ne 0$. 
In the deconfined theory, 
the following symmetry breaking occur: 
\bea
SO(N) + (N+6) \fund &\to& SO(N-1) + (N+5) \fund, \\
SO(N+5) + (N+1) \fund &\to& SO(N+5) + (N-1) \fund
\eea
The low energy deformed deconfined theory is given 
in Table 7, 
\begin{table}[h]
\begin{center}
\begin{tabular}{c|cc}
& $SO(N-1)$ & $SO(N+5)$ \\
\hline
$y'$ & $\fund$ & $\fund$ \\ 
$s$ & ${\bf 1}$ & ${\bf 1}$ \\
\end{tabular}
\end{center}
\label{ddec1}
\caption[dual]{The deformed deconfined theory 
for $\langle p \rangle$}
\end{table}
and the effective superpotential is 
\be
W_{{\rm eff}}= m(y'^2)_{{\rm singlet}}^2. 
\ee
In the dual, the direction under consideration 
corresponds to $\langle (p^2) \rangle \ne 0$. 
Then, the low energy deformed dual theory is 
displayed in Table 8, 
\begin{table}[h]
\begin{center}
\begin{tabular}{c|cc}
& $SO(10)$ & $SO(N+5)$ \\
\hline
$\tilde{y}$ & $\fund$ & $\fund$ \\ 
$s$ & ${\bf 1}$ & ${\bf 1}$ \\
$(y^2)$ & ${\bf 1}$ & $\sym \oplus {\bf 1}$ \\
\end{tabular}
\end{center}
\label{ddual2}
\caption[dual]{The deformed dual theory 
for $\langle p \rangle$}
\end{table}
and the dual superpotential is 
\be
\label{spp}
\tilde{W}_{{\rm eff}}=(y^2) \tilde{y}^2 - 
\frac{1}{2m} \tilde{y}^4. 
\ee
This resulting theory is also consistent 
under the deformation along $\langle p \rangle \ne 0$. 
In fact, taking a dual of $SO(N-1)+(N+5)\fund$ \cite{IS} 
and integrating out the massive modes, 
we can easily derive the dual in Table 8 
and the superpotential (\ref{spp}).

Furthermore, we can check the consistency 
under the mass term deformation. 
%
%
%
%
%
%
Adding the mass term $\delta W= \frac{1}{2} m' p^2$ to 
the superpotential in the deconfined theory, 
and integrating out $p$, 
we can derive the effective theory: 
$SO(N)$ gauge theory with $(N+5)$ vectors and 
$SO(N+5)$ gauge theory with $(N+1)$ vectors 
and a singlet $s$. 
The effective superpotential takes the form 
\be
W_{{\rm eff}}=-\frac{1}{2m'}(yz)^2 + z^2 s + 
m (y^2)_{{\rm singlet}}^2.
\ee
On the dual side, this deformation corresponds to 
adding the term $\delta \tilde{W}=\frac{1}{2} m' (p^2)$ 
to the dual superpotential. 
The equation of motion for $(p^2)$ forces $\tilde{p}$ 
to develop a VEV. 
This leads to break $SO(10)$ to $SO(9)$, 
then we arrive at the following effective theories: 
$SO(9)$ gauge theory with $(N+5)$ vectors 
and $SO(N+5)$ gauge theory 
with a symmetric tensor and 10 vectors 
and a singlet $s$, a gauge singlet meson $(y^2)$.  
The effective dual superpotential becomes  
\bea
\tilde{W}_{{\rm eff}} &=& -\frac{1}{2} m' \tilde{y_0}^2 s 
+ (y^2) \tilde{y}^2 + (y^2) \tilde{y_0}^2 \nonumber \\ 
&-& \frac{1}{2m} (\tilde{y}^2)^2_{{\rm singlet}} - 
\frac{1}{2m} (\tilde{y_0}^2)^2_{{\rm singlet}} - 
\frac{1}{m} (\tilde{y}^2)_{{\rm singlet}}
(\tilde{y_0}^2)_{{\rm singlet}}, 
\eea
where $\tilde{y_0}$ stands for a field transformed as 
(${\bf 1}, \fund$) under $SO(9) \times SO(N+5)$, 
which should be identified with the field $z$ 
in the deconfined theory. 
By rescaling the fields appropriately,
one can see that the above result is consistent.




Although we have constructed 
the dual description for $SO(N)$ SUSY gauge theory 
with a symmetric traceless tensor 
by deconfinement technique, 
it is not so trivial to see that 
the theory under consideration has a non-trivial 
infrared fixed point at the origin of the moduli space 
since the gauge groups are products. 
Following Terning's argument \cite{Terning}, 
we would like to show explicitly 
that the theory has indeed the infrared fixed point.

We note that one can analyze the theory 
for an arbitrary ratio of the two dynamical scales 
$\Lambda_1, \Lambda_2$ thanks to holomorphy \cite{SW}, 
where $\Lambda_1, \Lambda_2$ are the scales of 
$SO(10)$ gauge theory with ($N+6$) flavors, 
$SO(N+5)$ gauge theory with a symmetric traceless tensor 
and ten vector flavors, respectively. 
Furthermore, there is no phase transition 
when the ratio is varied. 
There are three cases 
to be considered. 
For $N < 4$, 
$SO(N + 5)$ theory is asymptotically non-free 
and $SO(10)$ theory is asymptotically free. 
This implies $\Lambda_1 \ll \Lambda_2$ and 
if we renormalize the gauge coupling of 
$SO(N + 5)$ theory $g$ at the scale $\Lambda_1$, 
then $g(\mu \sim \Lambda_1) \ll 1$. 
For $4 < N \le 18$, 
$SO(N+5)$ theory becomes asymptotic free and 
the limit $\Lambda_1 \gg \Lambda_2$ corresponds to 
weak coupling of $SO(N + 5)$ theory. 
For $N > 18$, $SO(10)$ becomes asymptotically non-free, 
which implies $\Lambda_1 \gg \Lambda_2$ and 
$g(\mu \sim \Lambda_1) \ll 1$. 
For $N=4$, since the gauge coupling $g$ does not run, 
we can take an arbitrary small coupling. 
In any cases, the gauge coupling $g \to 0$ 
as the ratio $\Lambda_1/\Lambda_2 \to 0$ or $\infty$, 
so we can perform the perturbative analysis for $g$.

Let us first consider the zero-th order case in $g$, 
$i.e.$ $SO(N + 5)$ dynamics is turned off. 
The dimensions of the gauge invariant operators 
have to satisfy the following constraints 
to be in unitary representations of 
the superconformal algebra \cite{Mack}: 
\bea
\label{constraints1}
D(\tilde{y}^2) &=& 2 + 2 \gamma_{\tilde{y}}(g=0) \ge 1, \nonumber \\
D(\tilde{y}\tilde{p}) &=& 2 + \gamma_{\tilde{y}}(g=0) + 
\gamma_{\tilde{p}}(g=0) \ge 1, \nonumber \\
D(\tilde{y}^{10}) &=& 10 + 10 \gamma_{\tilde{y}}(g=0) \ge 1, \\
D(\tilde{y}\tilde{p}^9) &=& 10 + 9 \gamma_{\tilde{y}}(g=0) + 
\gamma_{\tilde{p}}(g=0) \ge 1, \nonumber \\
D((y^2)) &=& 1 + \gamma_{(y^2)}(g=0) \ge 1, \nonumber
\eea
where $\gamma_{\phi}$ is the anomalous dimension of 
the field $\phi$, 
and the bound is saturated for free fields. 
We note that the first term 
in the dual superpotential (\ref{finalsp})
is a product of three gauge invariant operators. 
Thus, these are irrelevant 
because they can be relevant 
only if the dimensions of these gauge invariants 
are one, which means that 
these operators are free. 
The fields $s$ interacts 
only through the first term which is irrelevant, 
so these are free fields and 
their anomalous dimensions vanish. 
Therefore the equalities (\ref{constraints1}) cannot 
be saturated.

In order to obtain more relations 
among the anomalous dimensions, 
we use the exact $\b$ function for 
the $SO(10)$ coupling $g_1$ \cite{NSVZ}
\be
\label{beta1}
\b(g_1) = -\frac{g_1^3}{16\pi^2} 
\frac{3 \times 8-(N+5)(1-\gamma_{\tilde{y}}(g=0)) 
- (1-\gamma_{\tilde{p}}(g=0))}{1-8\frac{g_1^2}{8\pi^2}},
\ee
and at the fixed point 
\be
\label{fp}
0 = 18 - N + (N+5)\gamma_{\tilde{y}}(g=0) + 
\gamma_{\tilde{p}}(g=0).
\ee
The second and the last term in Eq. (\ref{finalsp}) are 
relevant operators with R-charge 2 for $g = 0$,
so the following conditions have to be satisfied,
\bea
\label{constraints2}
D(\tilde{y}^4) &=& 4 + 4 \gamma_{\tilde{y}}(g=0)=3, \nonumber \\
D((y^2)\tilde{y}^2) &=& 3 + \g_{(y^2)}(g=0) + 
2 \g_{\tilde{y}}(g=0) = 3. 
\eea
On the other hand, 
$\tilde{y}^2, \tilde{p}^2$ and $s$ are gauge invariant 
operators for arbitrary $g$, 
and the corresponding constraint for the dimensions are 
\bea
\label{constraints3}
D(\tilde{y}^2) &=& 2 + 2 \g_{\tilde{y}}(g) \ge 1, \nonumber \\
D(\tilde{p}^2) &=& 2 + 2 \g_{\tilde{p}}(g) \ge 1, \\
D(s) &=& 2 + 2 \g_s(g) \ge 1. \nonumber
\eea
The second and the fourth inequalities of 
(\ref{constraints1}) and the first one of 
(\ref{constraints2}) lead to 
\be
\label{bound}
\g_{\tilde{p}}(g=0) > -\frac{3}{4},
\ee
therefore one can derive the bound for $N$ 
using the conditions (\ref{fp}), 
the first inequality of (\ref{constraints2}) and 
(\ref{bound}):
%
%
%
\be
\label{nbound}
N > \frac{64}{5}.
\ee
%
%
%
%
The above result implies that 
$SO(10)$ theory has an infrared fixed point 
if $N$ is in the range of (\ref{nbound}). 
Next, we would like to show 
that $SO(N+5)$ theory also has an infrared fixed point.

The exact beta function for $g$ is 
\be
\label{beta2}
\b(g) = -\frac{g^3}{16\pi^2} 
\frac{3 (N + 3) - 10 ( 1 - \g_{\tilde{y}}(g)) - 
(N+7)(1-\g_{(y^2)}(g))}{1 - (N+3) \frac{g^2}{8 \pi^2}}, 
\ee
where we assume that $\g_{\tilde{y}}$ and $\g_{(y^2)}$ 
can be expanded in $g$ perturbatively as follows 
\bea
\g_{\tilde{y}} &=& -\frac{g^2}{8\pi^2} \frac{N+4}{2} + 
{\cal O}(g^4), \\
\g_{(y^2)} &=& -\frac{g^2}{8\pi^2}(N+5) 
+ {\cal O}(g^4).
\eea
If the 1-loop beta function coefficient is negative 
but the 2-loop one is positive, 
then the infrared fixed point will exist \cite{BZ}:
\bea
\label{fp1}
\b_0 &=& - (2N - 8) - 10 \g_{\tilde{y}}(g=0) - 
(N+7) \g_{(y^2)}(g=0) < 0 \nonumber \\ 
&\Leftrightarrow& N > \frac{14}{5}, \nonumber \\
\b_1 &=& 5N + 20 +N^2 + 12N + 35 -(2N^2 - 2N -24) \\
&& - 10 (N+3) \g_{\tilde{y}}(g=0) - (N^2 + 10N + 21) 
\g_{(y^2)}(g=0) > 0 \nonumber \\ 
&\Leftrightarrow& -3 \le N \le 14 \nonumber, 
\eea
where $\b_0, \b_1$ denote 1- or 2-loop beta 
function coefficients.
Taking into account the conditions 
(\ref{nbound}), (\ref{fp1}), 
we find $N=13,14$. 




In summary, 
we have discussed 
a $\caln=1$ SUSY $SO(N)$ gauge theory with 
a symmetric traceless tensor. 
This theory saturates 
't Hooft matching conditions 
among the fundamental fields and 
the gauge invariant composites 
at the origin of the moduli space. 
This naively suggests a confining phase 
but Brodie, Cho, and Intriligator have conjectured 
that the origin of the moduli space is 
in a Non-Abelian Coulomb phase. 
If this is the case, 
it is natural to ask 
whether the dual description exists. 
In this paper, 
we have constructed its dual by the deconfinement 
technique. 
Since this approach leads to the product gauge groups, 
it is not so trivial that the theory has 
a non-trivial infrared fixed point. 
Following \cite{Terning}, 
we have shown that 
the theory has indeed a non-trivial fixed point 
at the origin of the moduli space for $N = 13, 14$. 
Thus, this result supports the argument of 
Brodie, Cho, and Intriligator. 
Of course, 
the dual description is not necessarily 
unique, 
we may be able to find other dualities 
by further explorations.

We hope this work will provide a useful guide 
to analyzing the theory 
where 't Hooft anomaly matching 
appears to be coincidental.

\vspace{2cm}

\begin{center}
{\large\bf Acknowledgements}
\end{center}
The author would like to thank S. Kitakado 
for a careful reading of the manuscript. 
This work is supported by Research Fellowships of the 
Japan Society for the Promotion of Science for 
Young Scientists (No. 3400).

\end{document}